\pdfoutput=1

\documentclass[12pt,a4paper]{article}

\usepackage{ifthen} 
\usepackage{booktabs}
\newboolean{pdflatex}
\setboolean{pdflatex}{true} 

\newboolean{articletitles}
\setboolean{articletitles}{true} 

\newboolean{uprightparticles}
\setboolean{uprightparticles}{false} 

\def\paperauthors{LHCb collaboration} 
\def\paperasciititle{Study of Z bosons produced in association with charm in the forward region} 
\def\papertitle{Study of \Z bosons produced \\ in association with charm \\ in the forward region} 
\def\paperkeywords{{High Energy Physics}, {LHCb}} 
\def\papercopyright{\the\year\ CERN for the benefit of the LHCb collaboration} 
\def\paperlicence{CC BY 4.0 licence}
\def\paperlicenceurl{https://creativecommons.org/licenses/by/4.0/}

\def\ztomm{\ensuremath{\Z\!\to\!\mu^+\mu^-}\xspace}
\def\drmuj{\ensuremath{\Delta R(\mu,j)}\xspace}
\def\ptj{\ensuremath{\pt(j)}\xspace}

\def\Zj{\ensuremath{\Z j}\xspace}
\def\Zc{\ensuremath{\Z c}\xspace}
\def\yZ{\ensuremath{y(\Z)}\xspace}
\def\Zcj{\ensuremath{\mathcal{R}^c\kern-0.4em{\raisebox{-0.2em}{$\scriptstyle j$}}}\xspace}
\def\mmm{\ensuremath{m(\mu^+\mu^-)}\xspace}
\def\intlumi{6\invfb}

\usepackage[top=1in, bottom=1.25in, left=1in, right=1in]{geometry}

\usepackage{microtype}
\usepackage{lineno}  
\usepackage{xspace} 
\usepackage{caption} 

\usepackage{graphicx}  
\usepackage{color}
\usepackage{colortbl}
\graphicspath{{./figs/}} 

\usepackage{amsmath} 
\usepackage{amssymb}
\usepackage{amsfonts}
\usepackage{upgreek} 

\newcommand*\patchAmsMathEnvironmentForLineno[1]{%
\expandafter\let\csname old#1\expandafter\endcsname\csname #1\endcsname
\expandafter\let\csname oldend#1\expandafter\endcsname\csname
end#1\endcsname
 \renewenvironment{#1}%
   {\linenomath\csname old#1\endcsname}%
   {\csname oldend#1\endcsname\endlinenomath}%
}
\newcommand*\patchBothAmsMathEnvironmentsForLineno[1]{%
  \patchAmsMathEnvironmentForLineno{#1}%
  \patchAmsMathEnvironmentForLineno{#1*}%
}
\AtBeginDocument{%
\patchBothAmsMathEnvironmentsForLineno{equation}%
\patchBothAmsMathEnvironmentsForLineno{align}%
\patchBothAmsMathEnvironmentsForLineno{flalign}%
\patchBothAmsMathEnvironmentsForLineno{alignat}%
\patchBothAmsMathEnvironmentsForLineno{gather}%
\patchBothAmsMathEnvironmentsForLineno{multline}%
\patchBothAmsMathEnvironmentsForLineno{eqnarray}%
}

\usepackage{hyperxmp}

\usepackage[pdftex,
            pdfauthor={\paperauthors},
            pdftitle={\paperasciititle},
            pdfkeywords={\paperkeywords},
            pdfcopyright={Copyright (C) \papercopyright},
            pdflicenseurl={\paperlicenceurl}]{hyperref}

\usepackage[bottom,flushmargin,hang,multiple]{footmisc}

\usepackage[all]{hypcap} 

\usepackage{xspace} 
\usepackage{upgreek}

\def\lhcb   {\mbox{LHCb}\xspace}

\def\MagUp {\mbox{\em Mag\kern -0.05em Up}\xspace}

\ifthenelse{\boolean{uprightparticles}}%
{

 \def\Ppi         {\ensuremath{\uppi}\xspace}

 \def\PDelta      {\ensuremath{\Delta}\xspace}                 
 \def\PXi         {\ensuremath{\Xi}\xspace}                 
 \def\PLambda     {\ensuremath{\Lambda}\xspace}                 
 \def\PSigma      {\ensuremath{\Sigma}\xspace}                 
 \def\POmega      {\ensuremath{\Omega}\xspace}                 
 \def\PUpsilon    {\ensuremath{\Upsilon}\xspace}

 \def\PB      {\ensuremath{\mathrm{B}}\xspace}                 
                  
 \def\PD      {\ensuremath{\mathrm{D}}\xspace}

 \def\PK      {\ensuremath{\mathrm{K}}\xspace}

 \def\PZ      {\ensuremath{\mathrm{Z}}\xspace}

 \def\Pi      {\ensuremath{\mathrm{i}}\xspace}

 \def\Ps      {\ensuremath{\mathrm{s}}\xspace}

 \def\thebaroffset{0.0em}
}
{

 \def\Ppi         {\ensuremath{\pi}\xspace}

 \mathchardef\PDelta="7101
 \mathchardef\PXi="7104
 \mathchardef\PLambda="7103
 \mathchardef\PSigma="7106
 \mathchardef\POmega="710A
 \mathchardef\PUpsilon="7107
                  
 \def\PB      {\ensuremath{B}\xspace}                 
                  
 \def\PD      {\ensuremath{D}\xspace}

 \def\PK      {\ensuremath{K}\xspace}

 \def\PZ      {\ensuremath{Z}\xspace}

 \def\Pi      {\ensuremath{i}\xspace}

 \def\Ps      {\ensuremath{s}\xspace}

 \def\thebaroffset{0.18em}
}
\newcommand{\offsetoverline}[2][\thebaroffset]{\kern #1\overline{\kern -#1 #2}}%

\makeatletter
\ifcase \@ptsize \relax
  \newcommand{\miniscule}{\@setfontsize\miniscule{4}{5}}
\or
  \newcommand{\miniscule}{\@setfontsize\miniscule{5}{6}}
\or
  \newcommand{\miniscule}{\@setfontsize\miniscule{5}{6}}
\fi
\makeatother

\DeclareRobustCommand{\optbar}[1]{\shortstack{{\miniscule (\rule[.5ex]{1.25em}{.18mm})}
  \\ [-.7ex] $#1$}}

\def\Z      {{\ensuremath{\PZ}}\xspace}

\def\squark    {{\ensuremath{\Ps}}\xspace}

\def\pion   {{\ensuremath{\Ppi}}\xspace}

\def\pip    {{\ensuremath{\pion^+}}\xspace}

\def\kaon    {{\ensuremath{\PK}}\xspace}

\def\KorKbar {\kern \thebaroffset\optbar{\kern -\thebaroffset \PK}{}\xspace}

\def\Km      {{\ensuremath{\kaon^-}}\xspace}

\def\D       {{\ensuremath{\PD}}\xspace}

\def\DorDbar {\kern \thebaroffset\optbar{\kern -\thebaroffset \PD}\xspace}
\def\Dz      {{\ensuremath{\D^0}}\xspace}

\def\Dp      {{\ensuremath{\D^+}}\xspace}
\def\Dm      {{\ensuremath{\D^-}}\xspace}

\def\DpDm    {\ensuremath{\Dp {\kern -0.16em \Dm}}\xspace}

\def\B       {{\ensuremath{\PB}}\xspace}

\def\BorBbar {\kern \thebaroffset\optbar{\kern -\thebaroffset \PB}\xspace}

\def\Bd      {{\ensuremath{\B^0}}\xspace}

\def\BdorBdbar {\kern \thebaroffset\optbar{\kern -\thebaroffset \Bd}\xspace}

\def\Bs      {{\ensuremath{\B^0_\squark}}\xspace}

\def\BsorBsbar {\kern \thebaroffset\optbar{\kern -\thebaroffset \Bs}\xspace}

\def\Y#1S{\ensuremath{\PUpsilon{(#1S)}}\xspace}

\def\LorLbar     {\kern \thebaroffset\optbar{\kern -\thebaroffset \PLambda}\xspace}

\def\to                 {\ensuremath{\rightarrow}\xspace}

\def\AT#1     {\ensuremath{A_{\mathrm{T}}^{#1}}\xspace}           

\def\C#1      {\ensuremath{\mathcal{C}_{#1}}\xspace}                       
\def\Cp#1     {\ensuremath{\mathcal{C}_{#1}^{'}}\xspace}                    
\def\Ceff#1   {\ensuremath{\mathcal{C}_{#1}^{\mathrm{(eff)}}}\xspace}        
\def\Cpeff#1  {\ensuremath{\mathcal{C}_{#1}^{'\mathrm{(eff)}}}\xspace}       
\def\Ope#1    {\ensuremath{\mathcal{O}_{#1}}\xspace}                       
\def\Opep#1   {\ensuremath{\mathcal{O}_{#1}^{'}}\xspace}                    

\newcommand{\aunit}[1]{\ensuremath{\text{\,#1}}}       

\newcommand{\tev}{\aunit{Te\kern -0.1em V}\xspace}
\newcommand{\gev}{\aunit{Ge\kern -0.1em V}\xspace}
\newcommand{\mev}{\aunit{Me\kern -0.1em V}\xspace}
\newcommand{\kev}{\aunit{ke\kern -0.1em V}\xspace}
\newcommand{\ev}{\aunit{e\kern -0.1em V}\xspace}
 
\newcommand{\mevc}{\ensuremath{\aunit{Me\kern -0.1em V\!/}c}\xspace}
\newcommand{\gevc}{\ensuremath{\aunit{Ge\kern -0.1em V\!/}c}\xspace}
\newcommand{\mevcc}{\ensuremath{\aunit{Me\kern -0.1em V\!/}c^2}\xspace}
\newcommand{\gevcc}{\ensuremath{\aunit{Ge\kern -0.1em V\!/}c^2}\xspace}

\def\fb   {\ensuremath{\aunit{fb}}\xspace}
\def\invfb   {\ensuremath{\fb^{-1}}\xspace}

\def\gsim{{~\raise.15em\hbox{$>$}\kern-.85em
          \lower.35em\hbox{$\sim$}~}\xspace}
\def\lsim{{~\raise.15em\hbox{$<$}\kern-.85em
          \lower.35em\hbox{$\sim$}~}\xspace}

\def\pt         {\ensuremath{p_{\mathrm{T}}}\xspace}

\def\evtgen     {\mbox{\textsc{EvtGen}}\xspace}

\def\geant      {\mbox{\textsc{Geant4}}\xspace}

\def\photos     {\mbox{\textsc{Photos}}\xspace}

\def\pythia     {\mbox{\textsc{Pythia}}\xspace}

\def\tell1  {TELL1\xspace}
\def\ukl1   {UKL1\xspace}

\usepackage{cite} 
\usepackage{mciteplus}

\usepackage{longtable} 

\begin{document}

\renewcommand{\thefootnote}{\fnsymbol{footnote}}
\setcounter{footnote}{1}

\begin{titlepage}
\pagenumbering{roman}

\vspace*{-1.5cm}
\centerline{\large EUROPEAN ORGANIZATION FOR NUCLEAR RESEARCH (CERN)}
\vspace*{1.5cm}
\noindent
\begin{tabular*}{\linewidth}{lc@{\extracolsep{\fill}}r@{\extracolsep{0pt}}}
\ifthenelse{\boolean{pdflatex}}
{\vspace*{-1.5cm}\mbox{\!\!\!\includegraphics[width=.14\textwidth]{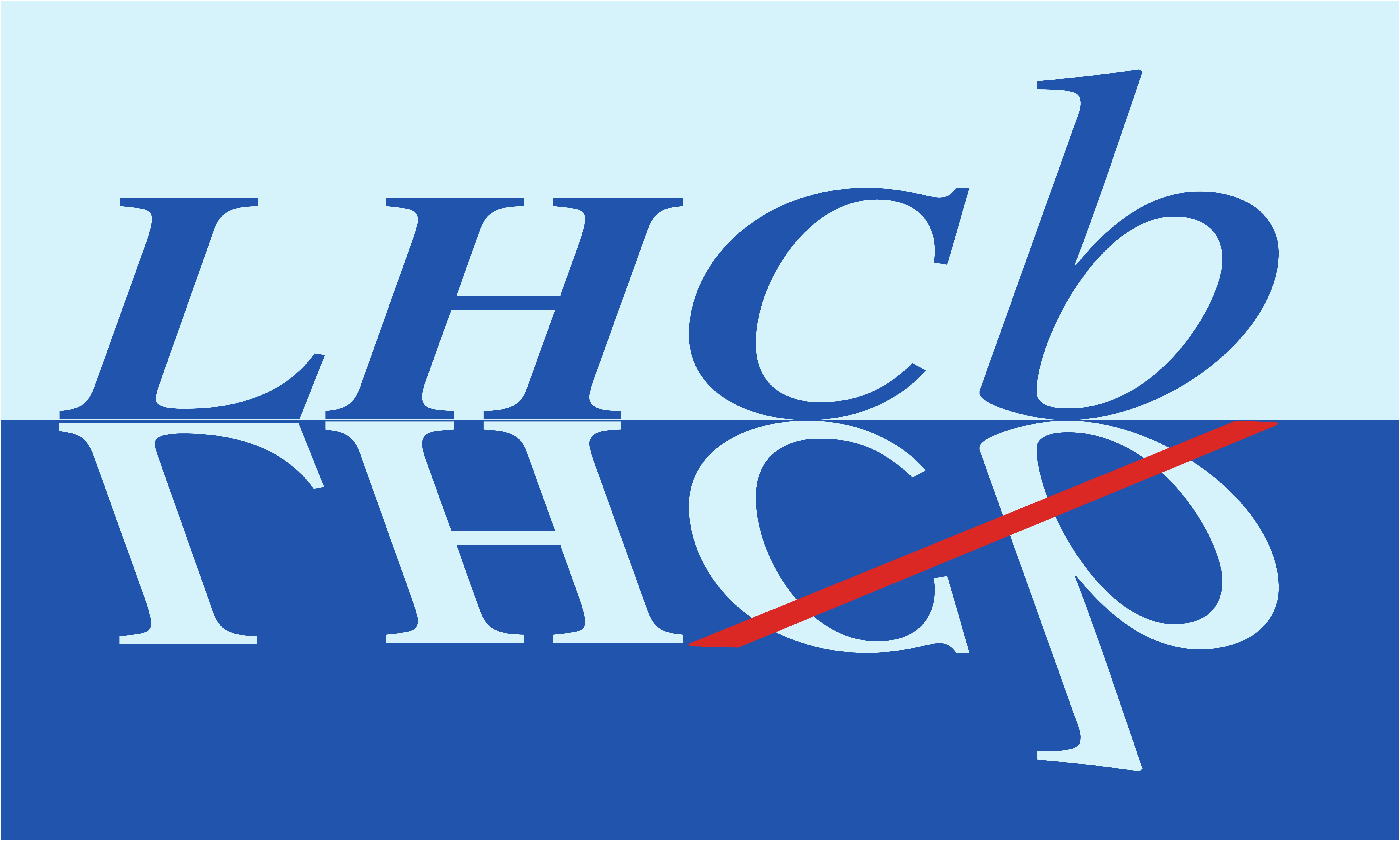}} & &}%
{\vspace*{-1.2cm}\mbox{\!\!\!\includegraphics[width=.12\textwidth]{lhcb-logo.eps}} & &}%
\\
 & & CERN-EP-2021-176 \\  
 & & LHCb-PAPER-2021-029 \\  
 & & \today \\
 & & \\
\end{tabular*}

\vspace*{4.0cm}

{\normalfont\bfseries\boldmath\huge
\begin{center}
  \papertitle
\end{center}
}

\vspace*{2.0cm}

\begin{center}

\paperauthors\footnote{Authors are listed at the end of this Letter.}
\end{center}

\vspace{\fill}

\begin{abstract}
  \noindent
  Events containing a \Z boson and a charm jet are studied for the first time in the forward region of proton-proton collisions.
  The data sample used corresponds to an integrated luminosity of \intlumi collected at a center-of-mass energy of $13\tev$ with the LHCb detector.
  In events with a \Z boson and a jet, the fraction of charm jets
  is determined in intervals of \Z-boson rapidity in the range $2.0 < \yZ < 4.5$.
  A sizable enhancement is observed in the forward-most \yZ interval, which could be indicative of a valence-like intrinsic-charm component in the proton wave function.
\end{abstract}

\vspace*{2.0cm}

\begin{center}
Published as Physical Review Letters {\bf 128} (2022) 082001
\end{center}

\vspace{\fill}

{\footnotesize
\centerline{\copyright~\papercopyright. \href{\paperlicenceurl}{\paperlicence}.}}
\vspace*{2mm}

\end{titlepage}

\newpage
\setcounter{page}{2}
\mbox{~}

\renewcommand{\thefootnote}{\arabic{footnote}}
\setcounter{footnote}{0}

\pagestyle{empty}
\cleardoublepage


\pagestyle{plain} 
\setcounter{page}{1}
\pagenumbering{arabic}


The possibility that the proton wave function may contain a $|uudc\bar{c}\rangle$ component,
referred to as intrinsic charm (IC),
in addition to the 
charm content that arises 
due to perturbative gluon radiation, {\em i.e.}\ $g\!\to\! c\bar{c}$ splitting, has been debated for decades (for a recent review, see Ref.~\cite{Brodsky:2015fna}).
The light front QCD calculations of Refs.~\cite{Brodsky:1980pb,Brodsky:1981se}, referred to as the BHPS model, predict that non-perturbative IC manifests as valence-like charm content in the parton distribution functions~(PDFs) of the proton; whereas, if the $c$-quark content is entirely perturbative in nature, the charm PDF resembles that of the gluon and sharply decreases at large momentum fractions, $x$. (Charge conjugation is implied throughout this Letter, {\em e.g.}, charm refers to both the $c$ and $\bar{c}$ quarks.)
Understanding the role that non-perturbative dynamics play inside the nucleon is a fundamental goal of nuclear physics~\cite{Hoffmann:1983ah,Navarra:1995rq,Franz:2000ee,Pumplin:2005yf,Freeman:2012ry,Broadsky:2012rw,Gong:2013vja,Hobbs:2013bia,Ball:2015tna,Blumlein:2015qcn,Duan:2016rkr,SUFIAN2020135633}. 
Furthermore, the existence of IC would have many phenomenological consequences. 
For example, IC would alter both the rate and kinematics of $c$ hadrons produced by cosmic-ray proton interactions in the atmosphere, which are 
an important source of background in studies of astrophysical neutrinos~\cite{Aartsen:2013jdh,Gauld:2015kvh,Halzen:2016thi,Laha:2016dri,Giannini:2018utr,Goncalves:2021yvw}.
The  cross sections of many processes at the LHC and other accelerators would also be affected~\cite{Lai:2007dq,Brodsky:2007yz,Stavreva:2009vi,Bednyakov:2013zta,Dulat:2013hea,Halzen:2013bqa,Rostami:2015iva,Boettcher:2015sqn,Bailas:2015jlc,Lipatov:2016feu,Bai:2018xum}.

Measurements of $c$-hadron production in deep inelastic scattering~\cite{Aubert:1982tt} and in fixed-target experiments~\cite{LHCb-PAPER-2018-023}, where the typical momentum transfers were $Q \lesssim 10\gev$ (natural units are used throughout this Letter), have been interpreted both as evidence for~\cite{Harris:1995jx,Steffens:1999hx} and against~\cite{Jimenez-Delgado:2014zga} the percent-level IC content predicted by BHPS.
Even though such experiments are in principle sensitive to valence-like $c$-quark content, interpreting these low-$Q$ data is challenging since it requires careful theoretical treatment of nonperturbative hadronic and nuclear effects. 
Recent global PDF analyses, which also include measurements from the LHC, are inconclusive and can only exclude IC carrying more than a few percent of the momentum of the proton~\cite{Hou:2017khm,Ball:2016neh}.

\begin{figure}[b]
\centering
\includegraphics[width=0.35\columnwidth]{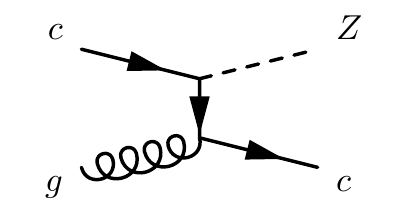}
\includegraphics[width=0.35\columnwidth]{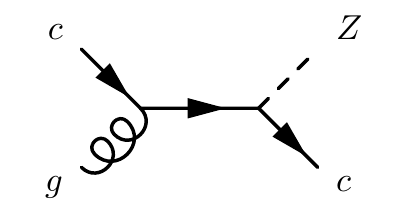}
\caption{Leading-order Feynman diagrams for $gc\!\to Zc$ production.}
\label{fig:diagrams}
\end{figure}

\begin{figure}[t]
\centering
\includegraphics[width=0.7\columnwidth]{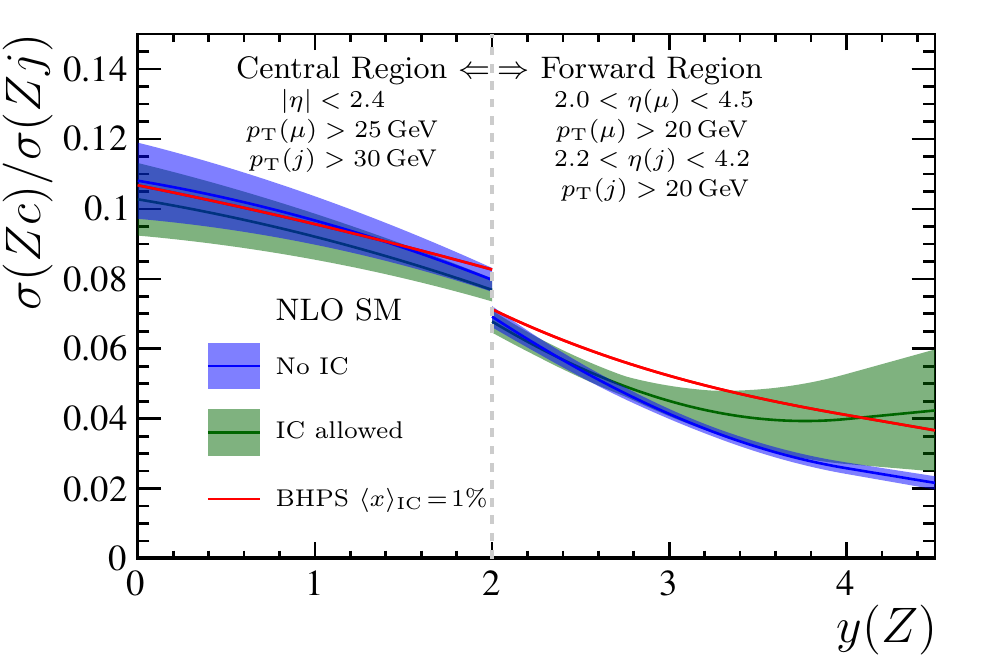}
\caption{NLO SM predictions~\cite{Boettcher:2015sqn} for \Zcj without IC~\cite{Butterworth:2015oua}, allowing for potential IC~\cite{Ball:2016neh}, and with the valence-like IC predicted by BHPS with a mean momentum fraction of 1\%~\cite{Hou:2017khm}. The fiducial region from Ref.~\cite{CMS:2020hmf} is used for $\yZ < 2$; otherwise the fiducial region of this analysis is employed. The broadening of the error band that arises in the forward region, when allowing for IC, is due to the lack of sensitivity to valence-like IC from previous experiments. More details on these calculations are provided in the Supplemental Material~\cite{Supp}.  The error bands shown for the first two predictions display the 68\% confidence-level regions. Only the central value is shown for BHPS due to the charm PDF being fixed.}
\label{fig:sensitivity}
\end{figure}

Reference~\cite{Boettcher:2015sqn} proposed probing IC by studying events containing a \Z boson and a charm jet, \Zc, 
in the forward region of proton-proton ($pp$) collisions at the LHC.
The ratio of production cross sections $\Zcj \equiv \sigma(\Zc)/\sigma(\Zj)$, where \Zj refers to events containing a \Z boson and any type of jet,  was chosen because it is less sensitive than $\sigma(\Zc)$  to experimental and theoretical uncertainties.
Since \Zc production is inherently at large $Q$, above the electroweak scale, hadronic effects are small. 
A leading-order \Zc production mechanism is $gc\!\to Zc$ scattering (see Fig.~\ref{fig:diagrams}), where in the forward region one of the initial partons must have large $x$, hence \Zc production probes the valence-like region (Fig.~\ref{fig:xdists} of the Supplemental Material shows the $x$ regions probed).
Using next-to-leading-order (NLO) Standard Model (SM) calculations, Fig.~\ref{fig:sensitivity} illustrates that a percent-level valence-like IC contribution would produce a clear enhancement in \Zcj for large (more forward) values of \Z rapidity, \yZ; whereas only small effects are expected in the central region where all previous measurements of \Zcj were made~\cite{D0:2013vwf,CMS:2020hmf}.

This Letter presents the first measurement of \Zcj in the forward region of $pp$ collisions.
The data sample used corresponds to an integrated luminosity of \intlumi collected at a center-of-mass energy of $\sqrt{s}=13\tev$ with the LHCb detector.
The \Z bosons are reconstructed using the \ztomm decay, where henceforth all $\Z/\gamma^* \to \mu^+\mu^-$ production in the mass range $60 < \mmm < 120\gev$ is labeled \ztomm.
The analysis is performed using jets clustered with the anti-$k_{\rm T}$ algorithm~\cite{antikt} using a distance parameter $R=0.5$. 
The fiducial region 
is defined in terms of the transverse momentum, \pt, pseudorapidity, $\eta$, and azimuthal angle, $\phi$, of the muon and jet momenta, and includes a requirement on
$\drmuj \equiv \sqrt{\Delta \eta(\mu,j)^2 + \Delta \phi(\mu,j)^2}$ to ensure that the muons and jets are well separated, which suppresses backgrounds from QCD multijet events and electroweak processes like $W+$jet production. 
Charm jets are the subset for which there is a promptly produced and weakly decaying $c$ hadron within the jet. 
The fiducial region is defined in Table~\ref{tab:fid}.
If multiple jets satisfy these criteria, the one with the highest \pt is selected.
No requirement is placed on the maximum number of jets in the event. 

\begin{table}[t]
  \begin{center}
    \caption{\label{tab:fid} Definition of the fiducial region.}
      \begin{tabular}{cc}
        \toprule
        \Z bosons & $\pt(\mu) > 20\gev$, $2.0 < \eta(\mu) < 4.5$, $60 < \mmm < 120\gev$ \\
        Jets & $20 < \ptj < 100\gev$, $2.2 < \eta(j) < 4.2$ \\
        Charm jets & $\pt(c$ hadron$) > 5\gev$, $\Delta R(j,c$ hadron$)<0.5$ \\
        Events & $\drmuj > 0.5$ \\
 \bottomrule
      \end{tabular}
  \end{center}
\end{table}

The quantity \Zcj is measured in intervals of \yZ as $\Zcj = N(c$-tag$) / [ \varepsilon(c$-tag$) N(j)]$, where $N(c$-tag$)$ is the observed \Zc yield, $\varepsilon(c$-tag$)$ is the $c$-tagging efficiency, and $N(j)$ is the total \Zj yield.
The integrated luminosity does not enter this expression because \Zcj involves a ratio of production cross sections.
In addition, the muon and jet reconstruction efficiencies largely cancel in the ratio due to the similarity of the \Z-boson and jet kinematics  in \Zc and \Zj production.
The $c$-tagging algorithm, which is described in detail in Ref.~\cite{ctag}, looks for a displaced-vertex (DV) signature inside the jet cone that is indicative of the weak decay of a $c$ hadron.

The \lhcb detector is a single-arm forward spectrometer covering the pseudorapidity range $2<\eta <5$, described in detail in Refs.~\cite{LHCb-DP-2008-001,LHCb-DP-2014-002}.
Simulated data samples are used to evaluate the detector response for jet reconstruction, including the $c$-tagging efficiency, and to validate the analysis.
In the simulation, $pp$ collisions are generated using
\pythia~\cite{Sjostrand:2007gs,*Sjostrand:2006za}
with a specific \lhcb configuration~\cite{LHCb-PROC-2010-056}.
Decays of unstable particles
are described by \evtgen~\cite{Lange:2001uf}, in which final-state
QED radiation is generated using \photos~\cite{davidson2015photos}.
The interaction of the generated particles with the detector, and its response,
are implemented using the \geant
toolkit~\cite{Allison:2006ve, *Agostinelli:2002hh} as described in
Ref.~\cite{LHCb-PROC-2011-006}.

The online event selection is performed by a trigger~\cite{LHCb-DP-2012-004,LHCb-DP-2019-001} consisting of a hardware stage using information from the calorimeter and muon systems, followed by a software stage that performs a full event
reconstruction.
At the hardware stage, events are required to have a muon with ${\pt(\mu) > 6\gev}$.
In the software stage, the muon track is required to be of good quality and to have ${\pt(\mu) > 10\gev}$.
The offline selection builds \ztomm candidates from two oppositely charged muon tracks that must be in the fiducial region defined in Table~\ref{tab:fid}  and consistent with originating directly from the same $pp$ collision.

Jet reconstruction is performed offline by clustering charged and neutral particle-flow candidates~\cite{LHCb-PAPER-2013-058} using the anti-$k_{\rm T}$ clustering algorithm as implemented in
\textsc{FastJet}~\cite{fastjet}.
Reconstructed jets with $15 < \ptj < 100\gev$ and $2.2 < \eta(j) < 4.2$ are kept for further analysis. Jets
with $15 < \ptj < 20\gev$, which are outside of the fiducial region,
are retained for use when unfolding the detector response.
The $\eta(j)$ requirement, which is included in the fiducial region  and was first used in Refs.~\cite{LHCb-PAPER-2015-016,LHCb-PAPER-2015-022,LHCb-PAPER-2015-021}, ensures a nearly uniform $c$-tagging efficiency of about 24\%,
with minimal \ptj or $\eta(j)$ dependence.
The fiducial requirement $\drmuj > 0.5$ is applied to reconstructed jets.
Finally, the highest-\pt jet satisfying these criteria from the same $pp$ collision as the \Z boson is selected.
After applying all requirements, 68\,694 \Zj candidates remain in the dataset. 

The effects of the detector response on the measured jet momenta are accounted for using an unfolding procedure.
This involves first determining the reconstructed \Zc and \Zj yields in intervals of $[\yZ,\ptj]$.
The non-\Z background is neglected for both \Zc and \Zj candidates 
because it is less than 1\% and largely cancels in the \Zcj ratio.
The $c$-jet yields are determined using the DV-based tagging approach described in detail in the following paragraphs.
Interval migration is accounted for by unfolding the $\ptj$ distributions of the \Zc and \Zj yields in each \yZ interval independently using
an iterative Bayesian procedure~\cite{D'Agostini:1994zf,Adye:2011gm}.
The \Zc yields are then corrected for $c$-tagging inefficiency. 
Finally, the unfolded $[\yZ,\ptj]$ distributions are integrated for $\ptj > 20\gev$ to obtain the \Zc and \Zj yields used to determine the \Zcj ratios. 
The analysis
employs three \yZ intervals with ranges 2.00--2.75, 2.75--3.50, and 3.50--4.50, and four \ptj intervals ranging 15--20, 20--30, 30--50, and 50--100\gev, where after unfolding the yields in the three highest \ptj intervals are summed to obtain \Zcj.

The signature of a $c$ jet is the presence of a long-lived $c$ hadron that carries a sizable fraction of the jet energy. 
The tagging of $c$ jets is performed using DVs formed from the decays of such $c$ hadrons.
The choice of using DVs and not single-track or other non-DV-based jet properties, {\em e.g.}\ the number of particles in the jet, is driven by the need for a small misidentification probability of light-parton jets.
Furthermore, the properties of DVs from $c$-hadron decays are known to be well modeled by simulation, which means that only small corrections using control samples are required.
Since DVs can also be formed from the decays of $b$ hadrons or due to artifacts of the reconstruction, the DV-tagged charm yields are obtained by fitting the distributions of DV features with good discrimination power between $c$, $b$, and light-parton jets. 

The tracks used as inputs to the DV-tagger algorithm are required to have $\pt > 0.5\gev$ and to be inconsistent with originating directly from a $pp$ interaction point. 
A DV is associated to a jet when $\Delta R < 0.5$ between the jet axis and the DV direction  of flight, defined by the vector from the $pp$ interaction point to the DV position. 
Requirements that reject strange-hadron decays and particles formed in interactions with material~\cite{LHCb-DP-2018-002} are placed on the mass, $m({\rm DV})$, and momentum, $p({\rm DV})$, of the particles that form the DV, along with the DV position.
In addition, 
only DVs with at most four tracks are used, since higher-multiplicity DVs are almost exclusively due to $b$-hadron decays.
More details about the $c$-tagging algorithm are provided in Ref.~\cite{ctag}.

\begin{figure}[t]
\centering
\includegraphics[width=0.49\columnwidth]{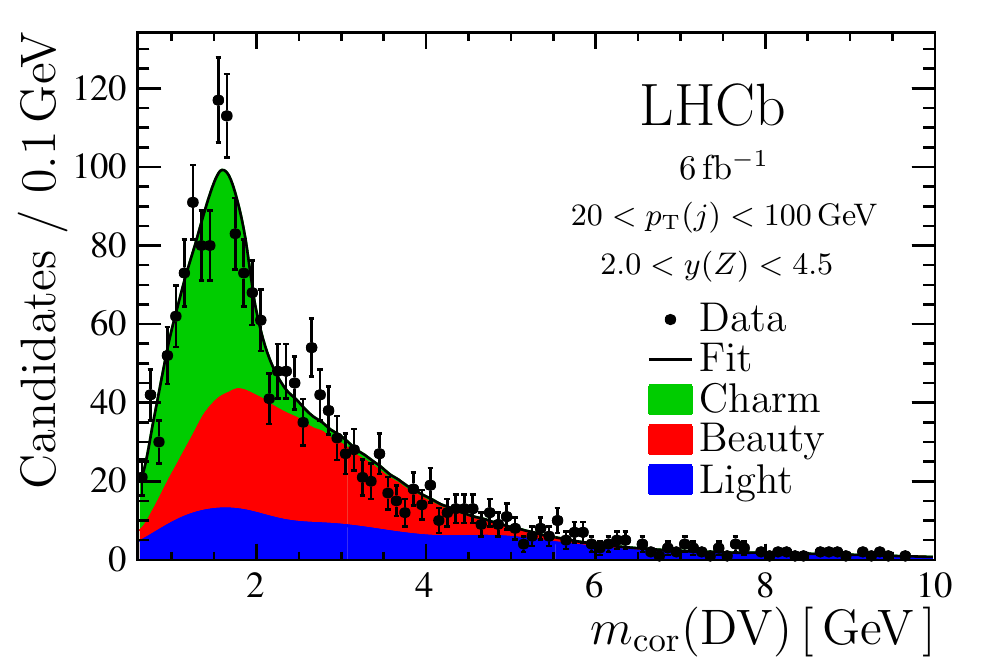}
\includegraphics[width=0.49\columnwidth]{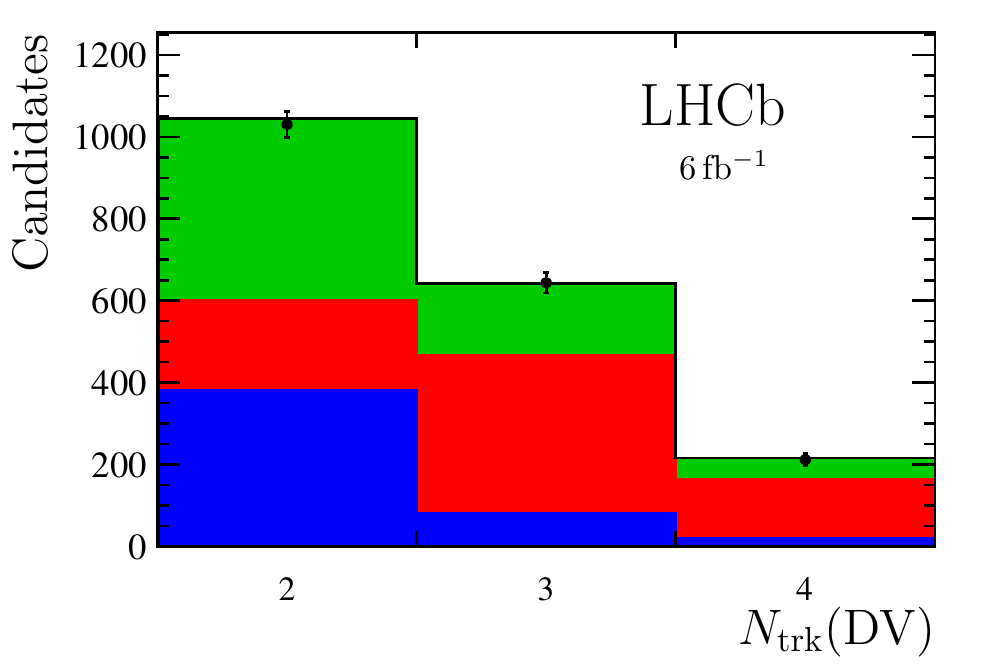}
\caption{Distributions of (left) $m_{\rm cor}({\rm DV})$ and (right) $N_{\rm trk}({\rm DV})$ for all DV-tagged candidates in the \Zj data sample reconstructed in the fiducial region with the projections of the fit results superimposed.}
\label{fig:svfits}
\end{figure}

Two DV properties are used to separate charm jets from beauty and light-parton jets: the number of tracks in the DV, $N_{\rm trk}({\rm DV})$, and the corrected mass, \mbox{$m_{\rm cor}({\rm DV}) \equiv \sqrt{m({\rm DV})^2 + [p({\rm DV})\sin{\theta}]^2} +  p({\rm DV})\sin{\theta}$}, where $\theta$ is the angle between the momentum and the flight direction of the DV.
The corrected mass, which is the minimum mass the long-lived hadron can have that is consistent with the flight direction, peaks near the typical $c$-hadron mass for $c$ jets, and consequently provides excellent discrimination against other jet types.
The DV track multiplicity provides additional discrimination against $b$ jets, since $b$-hadron decays often produce many displaced tracks.
These two distributions are fitted simultaneously to obtain the DV-tagged $c$-jet yields.
The probability density functions, referred to as templates, for $c$, $b$, and light-parton jets are obtained from calibration data samples that are each highly enriched in a given jet flavor~\cite{ctag}.
Figure~\ref{fig:svfits} shows the $m_{\rm cor}({\rm DV})$ and $N_{\rm trk}({\rm DV})$ distributions for all DV-tagged candidates in the \Zj data sample reconstructed in the fiducial region, along with the fit projections; such fits are performed in each $[\yZ,\ptj]$ interval to obtain the reconstructed \Zc yields.

The effects of $\ptj$ interval migration 
are corrected for using  the unfolding procedure. 
The detector response is studied using the \pt-balance distribution $\ptj/\pt(Z)$ for \Zj candidates that are nearly back-to-back in the transverse plane, using the same technique as in Refs.~\cite{LHCb-PAPER-2013-058,LHCb-PAPER-2016-064}.
Small adjustments are applied to the \ptj scale and resolution in simulation to obtain the best agreement with data.
In addition, for the \Zc and \Zj samples the \ptj and $\pt({\rm DV})$ distributions in simulation are adjusted to match those observed in data.
The unfolding matrix for jets that contain a reconstructed DV is shown in Fig.~\ref{fig:unfolding}, while the corresponding matrix for inclusive \Zj production
is provided in the Supplemental Material~\cite{Supp}.

\begin{figure}[t]
\centering
\includegraphics[width=0.58\columnwidth]{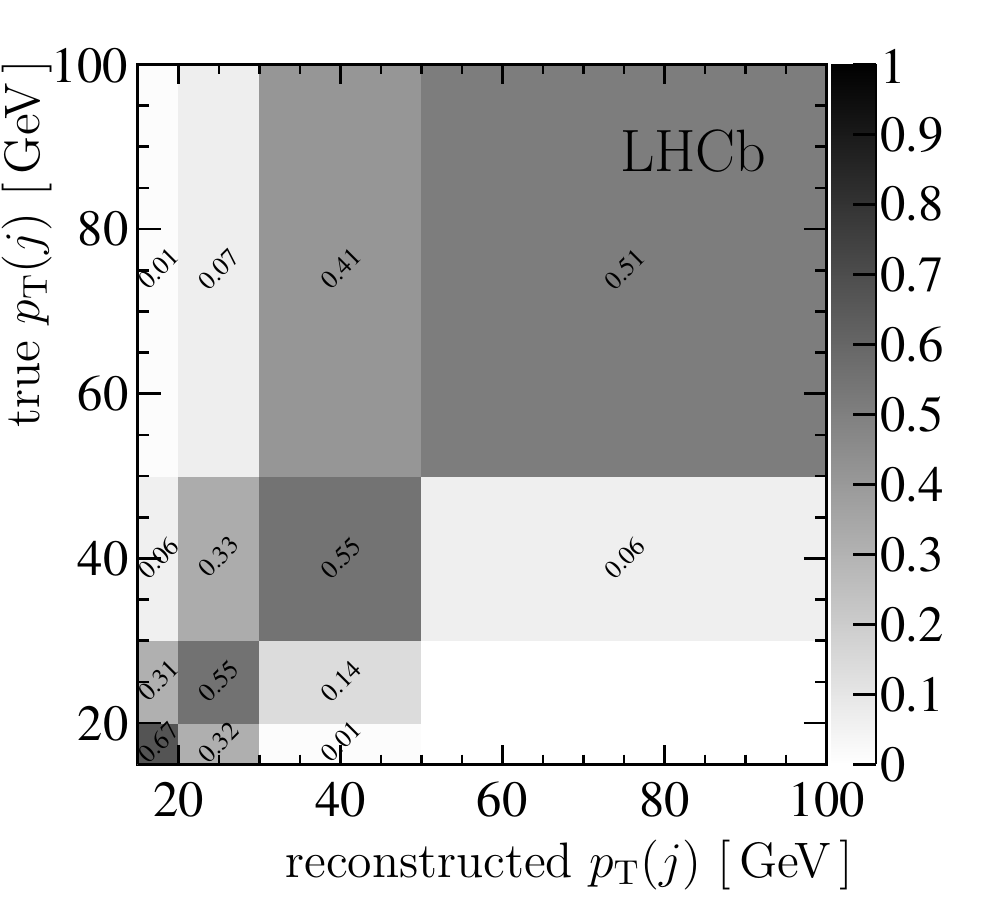}
\caption{The detector-response matrix for $c$-tagged jets. The shading represents the interval-to-interval migration probabilities ranging from (white) 0 to (black) 1. Numerical labels are only shown for values greater than 1\%.
Jets with true (reconstructed) \ptj in the 20--100\gev region but for which the reconstructed (true) \ptj is either below 15\gev or above 100\gev are included in the unfolding but not shown graphically.}
\label{fig:unfolding}
\end{figure}

The dominant systematic uncertainty is due to limited knowledge of the $c$-tagging efficiency, which is measured in \ptj intervals using data in Ref.~\cite{ctag} and briefly summarized here.
Scale factors that correct for discrepancies between data and simulation are determined using a  tag-and-probe approach on a dijet calibration sample.
A stringent requirement is applied to the tag jet which enriches the probe-jet sample in charm content.
The DV-tagged $c$-jet yield in the probe sample is obtained in the same way the \Zc yield is determined in this analysis, namely by fitting the $m_{\rm cor}({\rm DV})$ and $N_{\rm trk}({\rm DV})$ distributions for DV-tagged probe jets. 
The total number of $c$ jets in the probe sample is obtained by fully reconstructing the $\Dz \to \Km \pip$ and $\Dp \to \Km \pip\pip$ decays, obtaining the prompt-charm yields by fitting the \D-meson mass and impact-parameter distributions, then correcting these yields for the detector response, decay branching fractions~\cite{PDG2020}, and $c$-hadron fragmentation fractions~\cite{Lisovyi:2015uqa}.
The $c$-tagging efficiency is the ratio of the DV-tagged and total $c$-jet probe-sample yields. 
The scale factors that correct the $c$-tagging efficiency in simulation are determined to be $1.03 \pm 0.06$, $1.01 \pm 0.08$, and $1.09 \pm 0.17$ in the 20--30, 30--50, and 50--100\gev \ptj intervals, respectively, with corresponding  $c$-tagging efficiencies of $(23.9\pm1.4)\%$, $(24.4\pm 1.9)\%$, and $(23.6\pm4.1)\%$. 
These uncertainties, which include all statistical and systematic contributions, are propagated to the \Zcj results producing 6--7\% relative uncertainties in each \yZ interval.

Other  sources of smaller systematic uncertainty are also considered.
First, variations of the $m_{\rm cor}({\rm DV})$ and $N_{\rm trk}({\rm DV})$ templates are studied, which arise from using different strategies to model the backgrounds in the highly enriched calibration data samples.  
However, the shifts observed in the \Zc yields largely cancel with the corresponding shifts seen in $\varepsilon(c$-tag$)$. 
The residual differences of 3--4\% in each \yZ interval are assigned as systematic uncertainties.
The ratio of the jet-reconstruction efficiency for $c$ and inclusive jets is consistent with unity in all kinematic intervals in simulation, with a 1\% systematic uncertainty assigned due to the limited sample sizes.
Finally, the statistical precision of the back-to-back \Zj sample 
used to determine the \ptj scale and resolution is propagated through the unfolding procedure resulting in a 1\% relative systematic uncertainty on \Zcj.
The systematic uncertainties are summarized in Table~\ref{tab:syst}.

\begin{table}[t]
  \begin{center}
    \caption{\label{tab:syst} Relative systematic uncertainties on \Zcj, where ranges indicate that the value depends on the \yZ intervals.}
      \begin{tabular}{lc}
        \toprule
        Source & Relative Uncertainty \\
        \midrule
        $c$ tagging & 6--7\% \\
        DV-fit templates & 3--4\% \\
        Jet reconstruction & 1\% \\
        Jet \pt scale \& resolution & 1\% \\
        \midrule
        Total & 8\% \\
 \bottomrule
      \end{tabular}
  \end{center}
\end{table}

\begin{figure}[t]
\centering
\includegraphics[width=0.7\columnwidth]{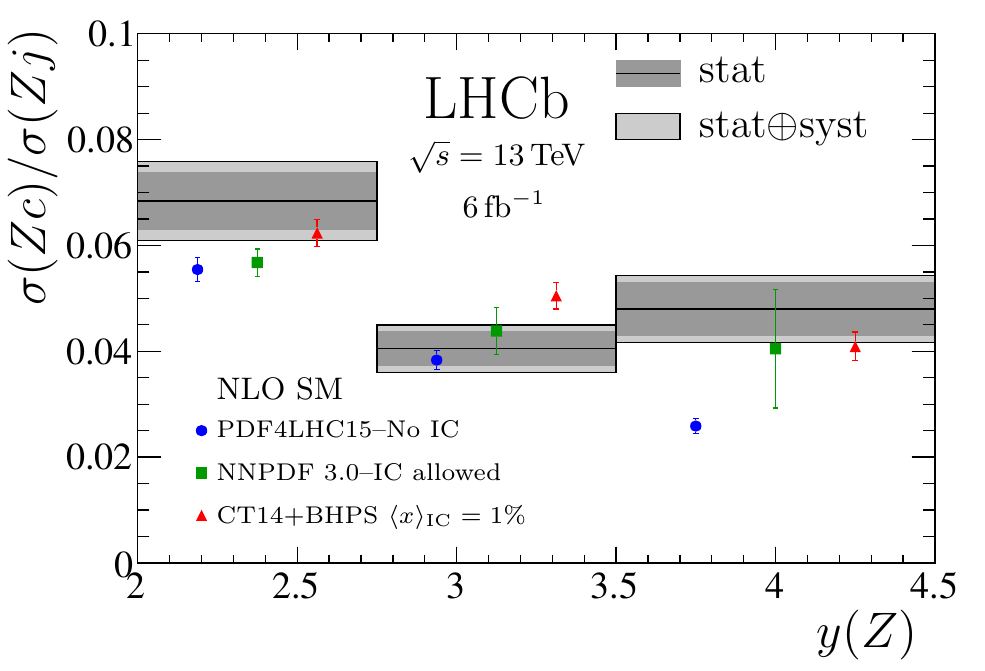}
\caption{Measured \Zcj distribution (gray bands) for three intervals of forward \Z rapidity, compared to NLO SM predictions~\cite{Boettcher:2015sqn} without IC~\cite{Butterworth:2015oua}, with the charm PDF shape allowed to vary (hence, permitting IC)~\cite{Ball:2016neh,NNPDF:2014otw}, and with IC as predicted by BHPS with a mean momentum fraction of 1\%~\cite{Hou:2017khm}. The predictions are offset in each interval to improve visibility.}
\label{fig:results}
\end{figure}

Figure~\ref{fig:results} shows the measured \Zcj distribution in intervals of \yZ; the numerical results are provided in Table~\ref{tab:zcj}, and additional results are reported in the Supplemental Material~\cite{Supp}. 
The measured \Zcj values are compared to NLO SM calculations~\cite{Boettcher:2015sqn} based on Refs.~\cite{Alioli:2010qp,Sjostrand:2014zea,Nason:2004rx,Alwall:2014hca,Frederix:2012ps,Golonka:2005pn,Giele:1998gw}, which are validated against additional predictions~\cite{Alwall:2014hca,Frederix:2012ps,Campbell:2011bn,Campbell:2015qma} and updated here to use more recent PDFs~\cite{Butterworth:2015oua,Hou:2017khm,Ball:2016neh,NNPDF:2014otw,Harland-Lang:2014zoa}. While $Zc$ predictions at NNLO in QCD are not available, $Zb$ predictions are~\cite{Gauld:2020deh}, and similar methods should be applicable.
The NNPDF analysis provides results where the charm PDF is allowed to vary, both in size and in shape~\cite{Ball:2016neh}.  
The sizable uncertainties that arise in the forward region are due to the lack of sensitivity to valence-like IC from previous experiments.
Reference~\cite{Hou:2017khm} updated the CT14 analysis~\cite{Dulat:2015mca} to include the IC content predicted by BHPS~\cite{Brodsky:1980pb,Brodsky:1981se}, which results in the enhancement at forward \yZ shown previously in Fig.~\ref{fig:sensitivity}.   
These predictions have smaller uncertainties because the size and shape of the IC contribution are fixed, {\em i.e.}\ the IC contribution is assumed to be known, hence does not contribute to the uncertainty on \Zcj. 
More details on the theory calculations, along with predictions based on other PDFs~\cite{Alekhin:2018pai,Jimenez-Delgado:2014twa,H1:2015ubc}, are provided in the Supplemental Material~\cite{Supp}. 

\begin{table}[t]
  \begin{center}
    \caption{\label{tab:zcj} Numerical results for the \Zcj measurements, where the first uncertainty is statistical and the second is systematic.}
      \begin{tabular}{lc}
        \toprule
        \yZ & \Zcj (\%) \\
        \midrule
        2.00--2.75 & $6.84\pm0.54\pm0.51$ \\
        2.75--3.50 & $4.05\pm0.32\pm0.31$ \\
        3.50--4.50 & $4.80\pm0.50\pm0.39$ \\
        \midrule
        2.00--4.50 & $4.98 \pm 0.25 \pm 0.35$ \\
 \bottomrule
      \end{tabular}
  \end{center}
\end{table}

The observed \Zcj values are consistent  with both the no-IC and IC hypotheses in the first two \yZ intervals; however, this is not the case in the forward-most interval where the ratio of the observed to no-IC-expected values is $1.85\pm0.25$. 
As illustrated in Fig.~\ref{fig:sensitivity}, this is precisely the \yZ region where valence-like IC would cause a large enhancement. 
Indeed, Fig.~\ref{fig:results} shows that, after including the IC PDF shape predicted by BHPS with a mean momentum fraction of 1\%, the theory predictions are consistent with the data. 
Incorporating these novel forward \Zcj results into a global analysis should strongly constrain the large-$x$ charm PDF, both in size and in shape. 
While the large enhancement in the forward-most \yZ interval is suggestive of valence-like IC, no definitive statements can be made until the \Zcj results are included in a global PDF analysis. 

In conclusion,
events containing a \Z boson and a charm jet 
are studied for the first time in the forward region of $pp$ collisions.
The data sample used corresponds to an integrated luminosity of \intlumi collected at a center-of-mass energy of $13\tev$ with the LHCb detector.
The ratio \Zcj 
is measured in intervals of \yZ and compared to NLO SM calculations.
The observed  spectrum exhibits a sizable enhancement at forward \Z rapidities, consistent with the effect expected if the proton wave function contains the $|uudc\bar{c}\rangle$ component predicted by BHPS. However, conclusions about whether the proton contains valence-like intrinsic charm can only be drawn after incorporating these results into global PDF analyses.

\section*{Acknowledgements}
%
%
\noindent We express our gratitude to our colleagues in the CERN
accelerator departments for the excellent performance of the LHC. We
thank the technical and administrative staff at the LHCb
institutes.
We acknowledge support from CERN and from the national agencies:
CAPES, CNPq, FAPERJ and FINEP (Brazil); 
MOST and NSFC (China); 
CNRS/IN2P3 (France); 
BMBF, DFG and MPG (Germany); 
INFN (Italy); 
NWO (Netherlands); 
MNiSW and NCN (Poland); 
MEN/IFA (Romania); 
MSHE (Russia); 
MICINN (Spain); 
SNSF and SER (Switzerland); 
NASU (Ukraine); 
STFC (United Kingdom); 
DOE NP and NSF (USA).
We acknowledge the computing resources that are provided by CERN, IN2P3
(France), KIT and DESY (Germany), INFN (Italy), SURF (Netherlands),
PIC (Spain), GridPP (United Kingdom), RRCKI and Yandex
LLC (Russia), CSCS (Switzerland), IFIN-HH (Romania), CBPF (Brazil),
PL-GRID (Poland) and NERSC (USA).
We are indebted to the communities behind the multiple open-source
software packages on which we depend.
Individual groups or members have received support from
ARC and ARDC (Australia);
AvH Foundation (Germany);
EPLANET, Marie Sk\l{}odowska-Curie Actions and ERC (European Union);
A*MIDEX, ANR, IPhU and Labex P2IO, and R\'{e}gion Auvergne-Rh\^{o}ne-Alpes (France);
Key Research Program of Frontier Sciences of CAS, CAS PIFI, CAS CCEPP, 
Fundamental Research Funds for the Central Universities, 
and Sci. \& Tech. Program of Guangzhou (China);
RFBR, RSF and Yandex LLC (Russia);
GVA, XuntaGal and GENCAT (Spain);
the Leverhulme Trust, the Royal Society
 and UKRI (United Kingdom).

\addcontentsline{toc}{section}{References}
\bibliographystyle{LHCb}
\bibliography{IC,main,standard,LHCb-PAPER,LHCb-CONF,LHCb-DP,LHCb-TDR}

\newpage
\clearpage
\newpage
\setcounter{equation}{0}
\setcounter{figure}{0}
\setcounter{table}{0}
\setcounter{section}{0}
\setcounter{page}{1}
\makeatletter
\renewcommand{\theequation}{S\arabic{equation}}
\renewcommand{\thefigure}{S\arabic{figure}}
\renewcommand{\thetable}{S\arabic{table}}
\renewcommand{\thepage}{S\arabic{page}}
\newcommand\ptwiddle[1]{\mathord{\mathop{#1}\limits^{\scriptscriptstyle(\sim)}}}

\begin{center}
\textbf{\Large Supplemental Material for LHCb-PAPER-2021-029} \\
\vspace{0.15in}
\textbf{\large \it Study of \Z bosons produced in association with charm in the forward region} \\
\end{center}

\subsection*{Theory Predictions}

The NLO SM calculations of Ref.~\cite{Boettcher:2015sqn} are updated here to use more recent PDFs~\cite{Butterworth:2015oua,Hou:2017khm,Ball:2016neh,NNPDF:2014otw,Harland-Lang:2014zoa,Alekhin:2018pai,Jimenez-Delgado:2014twa,H1:2015ubc}. 
These  predictions are made using the $Zj$ \textsc{PowhegBox} matrix element~\cite{Alioli:2010qp} interfaced to the \textsc{Pythia~8} parton shower~\cite{Sjostrand:2014zea} using \textsc{Powheg} matching~\cite{Nason:2004rx}. These predictions were cross-checked against results produced with the aMC@NLO matrix element generator~\cite{Alwall:2014hca} interfaced to the \textsc{Pythia~8} parton shower using \textsc{FxFx} matching~\cite{Frederix:2012ps}. Hadronization was performed with the \textsc{Pythia~8} event generator, while hadrons were decayed with the \textsc{EvtGen} package~\cite{Lange:2001uf} interfaced with the \textsc{Photos} final state radiation generator~\cite{Golonka:2005pn}. No requirement is placed on the maximum number of jets in the event. The results in Ref.~\cite{Boettcher:2015sqn} were produced using the CT14 NNLO PDF set~\cite{Dulat:2015mca} for the matrix element, and the corresponding LO PDF set for the parton shower. Using the kinematics of the initiating partons, these results were then weighted in this analysis to produce predictions from more recent PDF sets.

The associated systematic uncertainty for these predictions consists of PDF, scale, and strong-coupling uncertainty. 
Varying the charm mass leads to negligible changes in the \Zcj predictions. 
Full correlations were considered when evaluating these uncertainties. The PDF uncertainty is determined using the technique of Monte Carlo PDF replicas~\cite{Giele:1998gw}, while the scale uncertainty is determined from the envelope obtained by independently varying both the factorization and renormalization scales by up to a factor of two. The strong coupling uncertainty is evaluated by varying $\alpha_s$ within experimental uncertainty. Both the scale and strong-coupling uncertainties largely cancel in the ratio, leaving the PDF as the dominant source of uncertainty.

These theory predictions are validated by comparing to the \Zcj results from the CMS collaboration at central \yZ~\cite{CMS:2020hmf}, where there is minimal sensitivity to IC.  
The CMS collaboration measured \Zcj to be $(10.2\pm0.9)\%$. 
Reference~\cite{CMS:2020hmf} also provides NLO SM predictions, $(11.1\pm0.3)\%$ and $(9.0\pm0.9)\%$, obtained using the aMC@NLO generator interfaced with the \textsc{Pythia~8} parton shower using \textsc{FxFx} matching and the fixed-order MCFM generator~\cite{Campbell:2011bn,Campbell:2015qma}, respectively. 
Repeating the procedure used to obtain the theory predictions presented in this Letter, but in the CMS fiducial region, gives $(9.6\pm1.0)\%$, $(9.5\pm1.0)\%$, and $(9.7\pm1.0)\%$ for the PDF4LHC15 no IC, NNPDF 3.0 IC allowed, and CT14 with BHPS PDFs, respectively. 
Therefore, the approach used here produces predictions that are consistent both with the CMS measurement and with the theory predictions provided in Ref.~\cite{CMS:2020hmf}. 

Finally, in Fig.~\ref{fig:results} the CT14 with BHPS results are from the PDF set labeled as BHPS3 in Ref.~\cite{Hou:2017khm}, which is provided for the value $\langle x \rangle_{\rm IC} = 1\%$.

\clearpage 

\subsection*{Additional Numerical Results}

Numerical results are provided in Table~\ref{tab:zcj}. The statistical uncertainties are uncorrelated between \yZ intervals, whereas the systematic uncertainties are approximately completely correlated. Since many of the systematic uncertainties are correlated between \yZ intervals, numerical results are also provided for the ratios of \Zcj values between pairs of intervals in Table~\ref{tab:zcjRatios}.

\begin{table}[h!]
  \begin{center}
    \caption{\label{tab:zcjRatios} Numerical results for the ratios $r^c_{j}(i/k) \equiv \Zcj[\yZ_i] / \Zcj[\yZ_k]$, where the first uncertainty is statistical and the second is systematic for each result. The labels low, mid, and high refer to \yZ ranges of 2.00--2.75, 2.75--3.50 and 3.50--4.50, respectively.}
      \begin{tabular}{lc}
        \toprule
        Ratio & Value \\
        \midrule
$r^c_{j}({\rm mid}/{\rm low})$ & $0.59\pm0.07\pm0.04$ \\
$r^c_{j}({\rm high}/{\rm low})$ & $0.70\pm0.09\pm0.05$ \\
$r^c_{j}({\rm high}/{\rm mid})$ & $1.19\pm0.16\pm0.05$ \\
 \bottomrule
      \end{tabular}
  \end{center}
\end{table}


\subsection*{Additional Figures}

\vspace{2em}

\begin{figure}[h!]
\centering
\includegraphics[width=0.7\columnwidth]{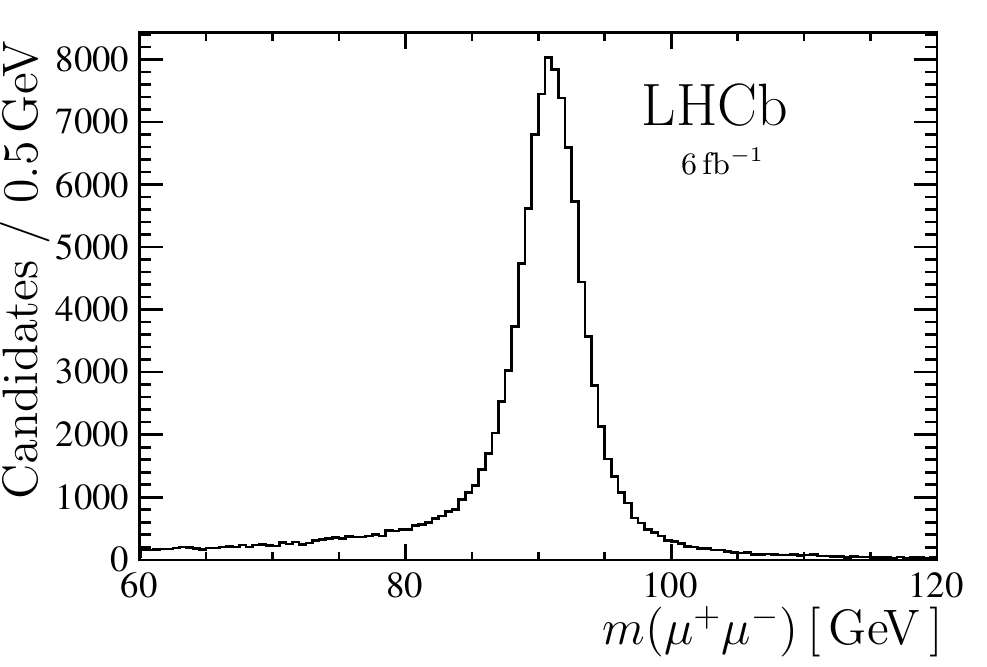}
\caption{Dimuon invariant mass distribution for the \Zj sample.}
\label{fig:m_mumu}
\end{figure}


\begin{figure}[h!]
\centering
\includegraphics[width=0.49\columnwidth]{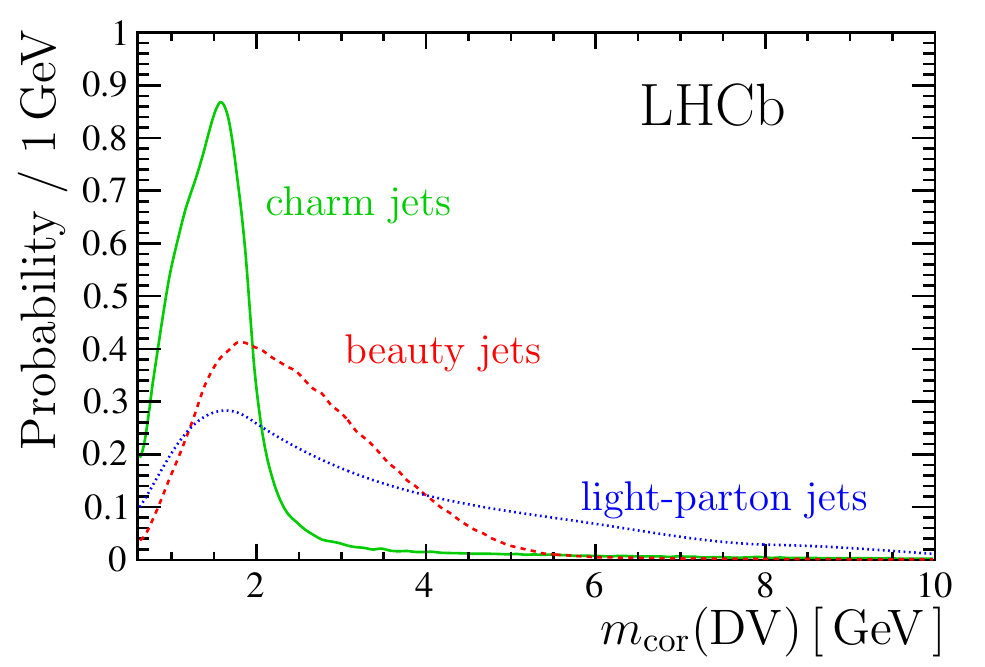}
\includegraphics[width=0.49\columnwidth]{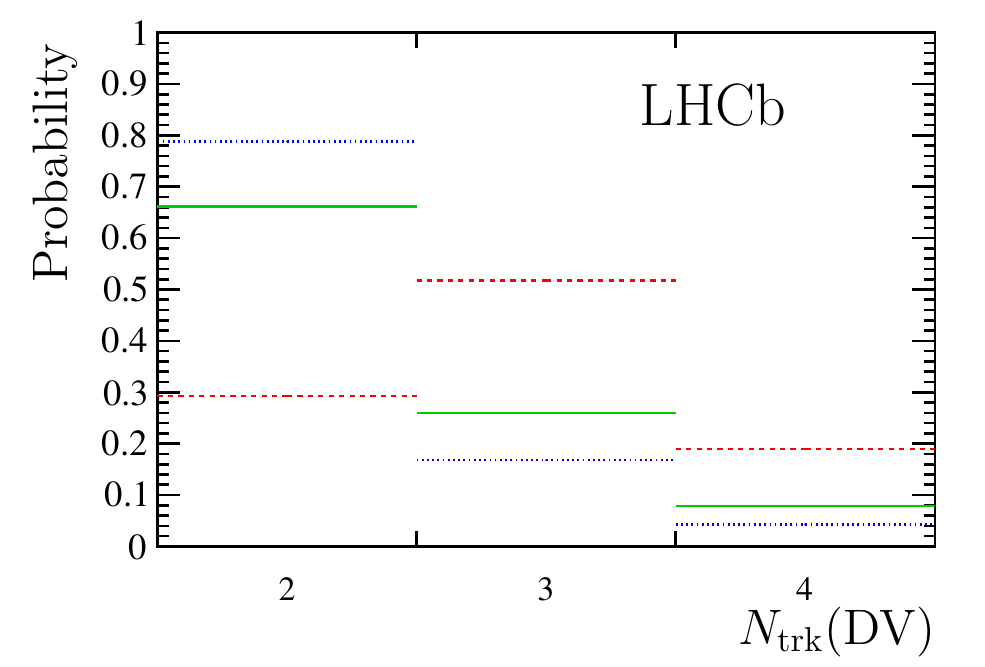}
\caption{Probability density functions for the DV features used in the $c$-tagging fits for $c$, $b$, and light-parton jets: (left) the corrected mass and (right) the track multiplicity. } 
\label{fig:svtemplates}
\end{figure}

\begin{figure}[h!]
\centering
\includegraphics[width=0.58\columnwidth]{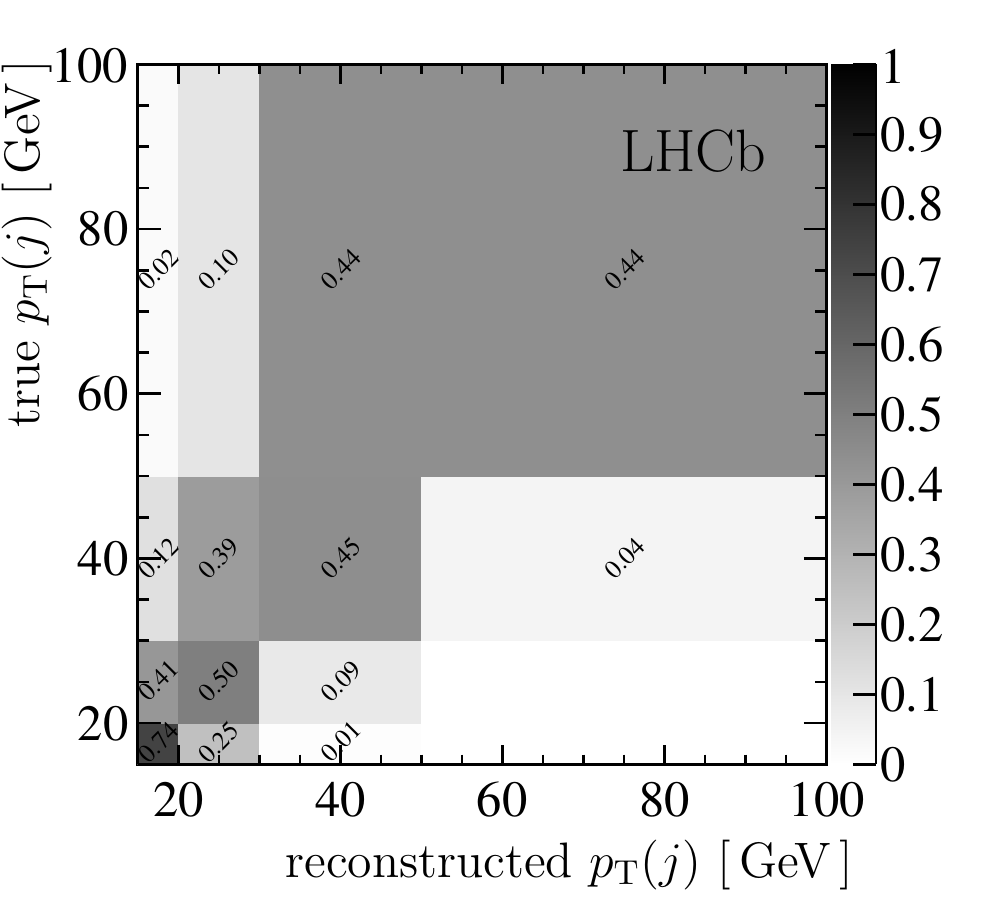}
\caption{
The detector-response matrix for inclusive \Zj events. 
The shading represents the interval-to-interval migration probabilities ranging from (white) 0 to (black) 1. Numerical labels are only shown for values greater than 1\%.
Jets with true (reconstructed) \ptj in the 20--100\gev region but for which the reconstructed (true) \ptj is either below 15\gev or above 100\gev are included in the unfolding but not shown graphically.}
\label{fig:unfolding2}
\end{figure}

\begin{figure}[h!]
\centering
\includegraphics[width=0.49\columnwidth]{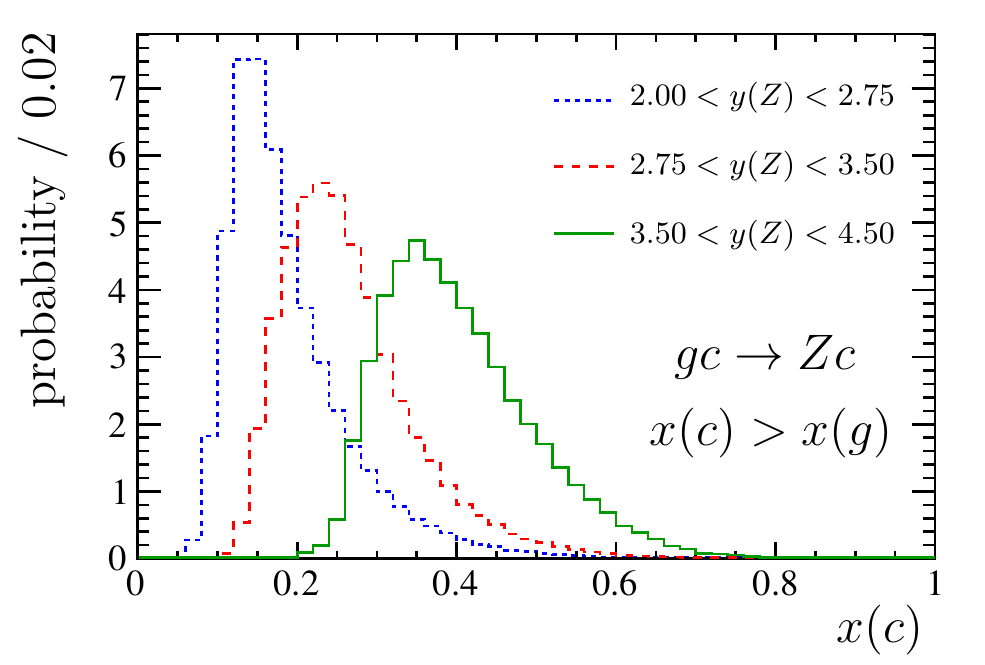}
\includegraphics[width=0.49\columnwidth]{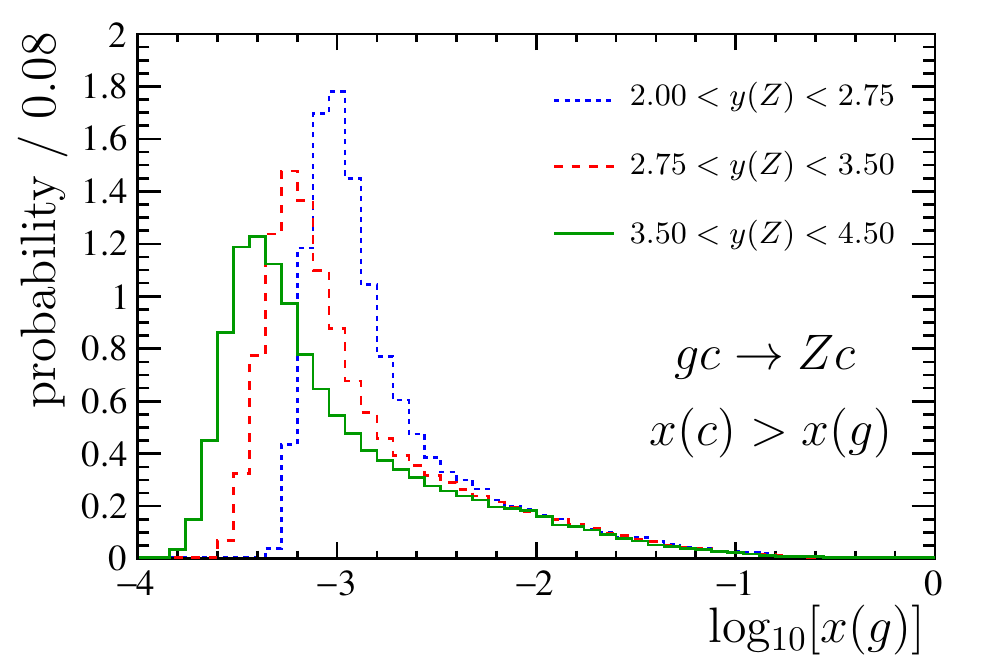}
\caption{Momentum fraction distributions of the (left) $c$ quark and (right) gluon for the process $g c \to Z c$ in the predominant scenario where the $c$ quark is the leading (higher-$x$) parton. Distributions are shown separately for the three $y(Z)$ intervals used in the analysis, with each distribution normalized to have an integral of unity. These distributions are obtained using the theory calculations described in detail here in the Supplemental Material.}
\label{fig:xdists}
\end{figure}

\begin{figure}[t]
\centering
\includegraphics[width=0.49\columnwidth]{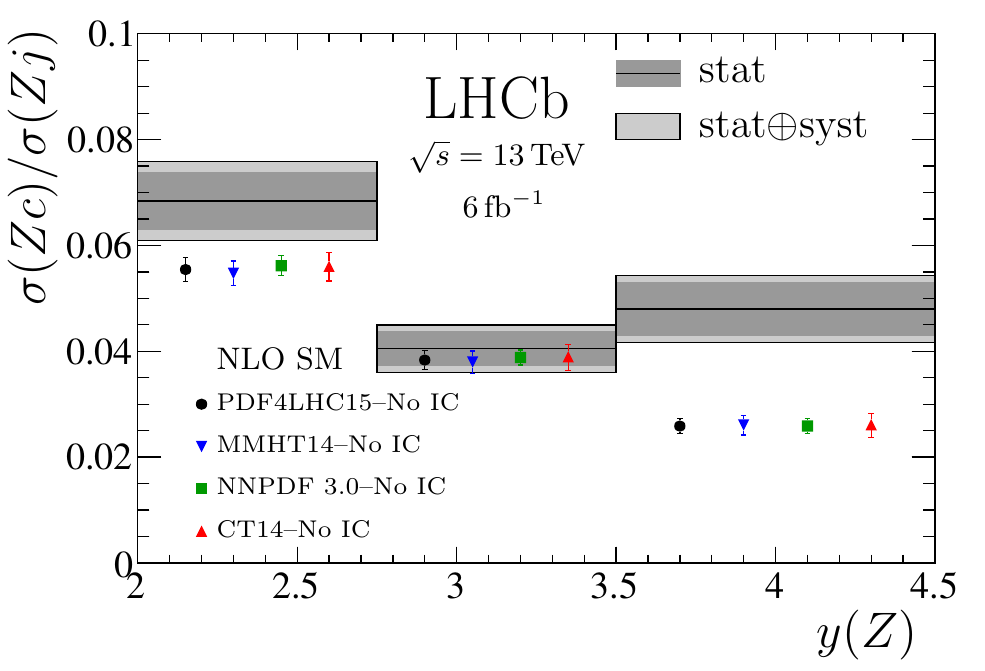}
\includegraphics[width=0.49\columnwidth]{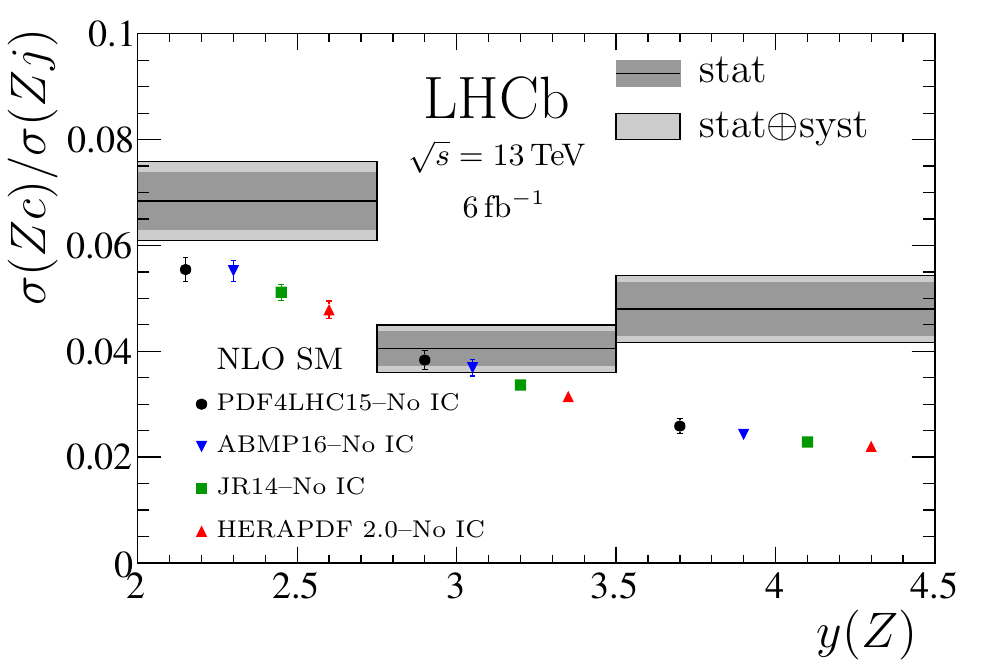}
\caption{Measured \Zcj distribution (gray bands) for three intervals of forward \Z rapidity, compared to NLO SM predictions~\cite{Boettcher:2015sqn} using various PDF sets. 
The predictions are offset in each interval to improve visibility.
The left plot shows that the three PDF sets~\cite{Dulat:2015mca,Harland-Lang:2014zoa,NNPDF:2014otw} on which the PDF4LHC15~\cite{Butterworth:2015oua} set is formed from all provide consistent predictions for \Zcj. The right plot shows that the ABM16~\cite{Alekhin:2018pai}, JR14~\cite{Jimenez-Delgado:2014twa}, and HERAPDF 2.0~\cite{H1:2015ubc} PDF sets also provide qualitatively similar predictions, though the JR14 and HERAPDF 2.0 predictions are shifted to lower \Zcj values.}
\label{fig:altpdfs}
\end{figure}

\clearpage

\setcounter{page}{1}
\pagenumbering{roman}

\newpage
\centerline
{\large\bf LHCb collaboration}
\begin
{flushleft}
\small
R.~Aaij$^{32}$,
A.S.W.~Abdelmotteleb$^{56}$,
C.~Abell{\'a}n~Beteta$^{50}$,
F.J.~Abudinen~Gallego$^{56}$,
T.~Ackernley$^{60}$,
B.~Adeva$^{46}$,
M.~Adinolfi$^{54}$,
H.~Afsharnia$^{9}$,
C.~Agapopoulou$^{13}$,
C.A.~Aidala$^{87}$
% ...

{\footnotesize \it

% ...
$^{9}$Universit{\'e} Clermont Auvergne, CNRS/IN2P3, LPC, Clermont-Ferrand, France\\
% ...
$^{13}$LPNHE, Sorbonne Universit{\'e}, Paris Diderot Sorbonne Paris Cit{\'e}, CNRS/IN2P3, Paris, France\\
% ...
$^{32}$Nikhef National Institute for Subatomic Physics, Amsterdam, Netherlands\\
% ...
$^{46}$Instituto Galego de F{\'\i}sica de Altas Enerx{\'\i}as (IGFAE), Universidade de Santiago de Compostela, Santiago de Compostela, Spain\\
% ...
$^{50}$Physik-Institut, Universit{\"a}t Z{\"u}rich, Z{\"u}rich, Switzerland\\
% ...
$^{54}$H.H. Wills Physics Laboratory, University of Bristol, Bristol, United Kingdom\\
% ...
$^{56}$Department of Physics, University of Warwick, Coventry, United Kingdom\\
% ...
$^{60}$Oliver Lodge Laboratory, University of Liverpool, Liverpool, United Kingdom\\
% ...
$^{68}$Syracuse University, Syracuse, NY, United States\\
% ...
$^{87}$University of Michigan, Ann Arbor, United States, associated to $^{68}$\\
\bigskip
% ...
}
\end{flushleft}

\end{document}